\documentclass[mathleft
]{an}
\usepackage{graphicx}
\usepackage{times}
\def\cobold{CO$^{\sf 5}$\-BOLD}
\begin{document}

% The following seven commands are intended for editorial usage and should be ignored by
% the author(s).
\Pagespan{789}{}% Document's page range. 
% If second parameter is left empty, the last page is computed automatically.
\Yearpublication{2007}%
\Yearsubmission{2006}%
\Month{11}%   
\Volume{999}%  
\Issue{88}% 
% \DOI{This.is/not.aDOI}% 

\title{First local helioseismic experiments with \cobold}

\author{O.~Steiner\inst{1}\fnmsep\thanks{Corresponding author:
\email{steiner@kis.uni-freiburg.de}\newline} \and G.~Vigeesh\inst{2} 
\and L.~Krieger\inst{1,2} \and S.~Wedemeyer-B\"ohm\inst{3}
\and W. Schaffenberger\inst{1} \and B. Freytag\inst{4}
}
\titlerunning{Helioseismic experiments with \cobold}
\authorrunning{O.~Steiner et al.}
\institute{
Kiepenheuer-Institut f\"ur Sonnenphysik, Sch\"oneckstrasse 6,
D-79104 Freiburg, Germany
\and 
Indian Institute of Astrophysics, Koramangala, Bangalore 560034, India
\and
Institute of Theoretical Astrophysics, University of Oslo, P.O.~Box 1029 Blindern,
N-0315 Oslo, Norway
\and
%Department of Physics and Astronomy, Michigan State University, East Lansing, MI 48824, U.S.A.
%\and
Centre de Recherche Astronomique de Lyon - Ecole Normale Sup\'erieure,
46, All\'ee d'Italie, F-69364 Lyon Cedex 07, France
}

\received{30 Dec 2006}
\accepted{30 Dec 2006}
\publonline{later}


\abstract{With numerical experiments we explore the feasibility of using
high frequency waves for probing the magnetic fields in the photosphere and the 
chromosphere of the Sun. We track a plane-parallel, monochromatic wave that
propagates through a non-stationary, realistic atmosphere, from the 
convection-zone through the photosphere into the magnetically dominated chromosphere,
where it gets refracted and reflected. We compare the wave travel time between two 
fixed geometrical height levels in the atmosphere (representing the formation
height of two spectral lines) with the topography of the surface of equal magnetic 
and thermal energy density (the magnetic canopy or $\beta=1$ contour) and find good 
correspondence between the two. We conclude that high frequency waves indeed bear 
information on the topography of the `magnetic canopy'.
}

\maketitle

\section{What is \cobold ?}

\cobold\ is a computer code designed for simulating hydrodynamic  processes including
radiative transfer in the ou\-ter and inner layers of stars. Additionally, it can treat the
interaction of magnetic fields with a plasma in the magnetohydrodynamic (MHD) approximation,
non-equilibrium chemical reaction networks, dynamic hydrogen ionization, and dust formation
in stellar atmospheres. \cobold\ stands for COnservative COde for the COmputation
of COmpressible COnvection in a BOx of L Dimensions with L=2,3.

In the past few years, \cobold\ has been extensively used for `box in a star' computations
(for a review see Steffen 2007), where a small piece of a star's interior and/or atmosphere
is simulated. 
A typical result of these kinds of computations is the emergent intensity from the stellar 
surface. In this way, the surface granulation pattern of the Sun and other stars, e.g., A-stars 
(Kochukhov et al.~2006) or red giants have been simulated. 
In the case of the Sun, a detailed comparison between high resolution observation
and simulation is possible and allows for a validation of the latter 
(Steffen, Ludwig, \& Freytag 2002).
On the other hand, \cobold\ is also used for `star in a box' computations in which
the computational box contains an entire star, e.g., the red supergiant Betelgeuse
(Freytag, Steffen, \& Dorch 2002; Chiavassa et al.~2006).

Recent applications of \cobold\ focus
on the dynamical evolution of carbon monoxide in the solar photosphere 
(Wedemeyer-B\"ohm et al. 2005; We\-demeyer-B\"ohm \& Steffen 2007) and on the
dynamical hydrogen ionization  (Leen\-aarts \& Wedemeyer-B\"ohm 2006), 
where deviations from statistical ionization equilibrium occur. 
These works have been carried out in
connection with efforts for a simulation including the layers from the 
convection zone to the chromosphere of the Sun (Wedemeyer et al.~2004), which
now have been expanded in order to include the effects of magnetic fields
(Schaffenberger et al.~2005, 2006). Details of these MHD-simulations are 
given in the next section.

\cobold\ works with Cartesian, non-equidistant but static grids,
realistic equations of state, multidimensional radiation transfer,
realistic opacities, and it features various boundary conditions.
As a simplest option the radiative transfer can be treated frequency-independently 
(`grey'). \linebreak Alternatively, the frequency-dependence can be taken into account by 
using a multi-group scheme that solves the radiative transfer problem 
for different opacity groups, which takes effects of line blanketing into account.

\begin{figure*}
  \centerline{\includegraphics[width=0.8\textwidth]{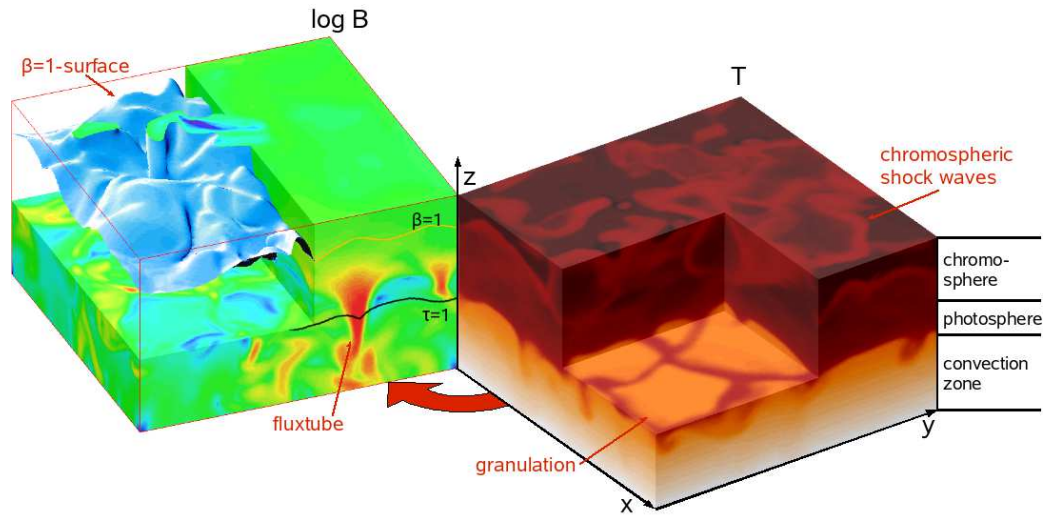}}
  \caption{{\bf Left:} Magnetic flux density on a color scale, where red 
  signifies strong (greater or approximately 0.1 T) and blue weak fields.
  The surface of equipartition between magnetic and thermal energy density 
  ($\beta = 1$) is indicated as blue surface (on the lateral boundaries by 
  a yellow curve). The black curve indicates optical depth unity. {\bf Right:}
  The temperature in colors shows granules and cool intergranular downflows
  at a height that corresponds to the mean optical depth unity. Note the
  mesh-work like pattern of hot shock waves in the chromospheric layers.
  From Wedemeyer-B\"ohm (2007).
  }
  \label{fig_3Dsection}
\end{figure*}

The multidimensional problem is reduced to 1-D problems by dimensional splitting.
Each of these 1-D problems is solved with a Godunov-type finite-volume scheme 
using an approximate Riemann solver mo\-di\-fied for a general equation of state and 
gravity. A Roe-type approximate Riemann solver is used for integration
of the hydrodynamic equations, whereas in the case of the MHD-equations
we use a HLL-solver of second order accuracy, instead. More details
on the MHD-solver can be found in Schaffenberger et al.~(2005). A good
introduction to the above mentioned numerical techniques can be
found in Laney (1998), LeVeque (2002), LeVeque et al.~(1998), or Toro (1999).
\cobold\ is programmed in FORTRAN 90 with OpenMP directives.
The manual for \cobold\ can be found under
{\sf www.astro.uu.se/\~{}bf/co5bold\_main.html}
while talks delivered at the first \cobold\ workshop are
at {\sf http://folk. uio.no/svenwe/cws2006/cws\_cont.html}.
For the analysis and visualization a \cobold\ analysis tool 
(CAT, {\sf http:// folk.uio.no/svenwe/co5bold/cat.html}) 
exists.

%%%%%%%%%%%%%%%%%%%%%%%%%%%%%%%%%%%%%%%%%%%%%%%%%%%%%

\section{MHD-simulation from the convection zone to the chromosphere}

A three-dimensional magnetohydrodynamic simulation of the 
integral layers from the upper convection zone to the middle chromosphere was
carried out by Schaffenberger et al. (2005, 2006). The computational domain
in this simulation extends over a height range of 2800 km, of which 1400 km reach 
below the mean surface of optical depth unity and 1400 km above it. The horizontal 
dimensions are $4800~{\rm km} \times 4800$~km.  With $120^3$ grid cells, the spatial 
resolution in the horizontal direction is 40~km, while in the vertical direction it is 
50~km through the convection-zone layer and 20~km throughout the photosphere and 
chromosphere. The lateral boundary conditions are periodic in 
all variables, whereas the lower boundary is `open' in the sense that the fluid can 
freely flow in and out of the computational domain under the condition of vanishing
total mass flux. The specific entropy of the inflowing mass is fixed to a
value previously determined so as to yield solar radiative flux at the
upper boundary. The upper boundary can be chosen to be transmitting so that
acoustic waves can leave the computational domain with no significant reflection 
at the boundary. Stress-free conditions are in effect for the horizontal velocities, 
viz., ${\rm d}v_{x,y}/{\rm d} z = 0$.

The MHD simulation starts with a homogeneous, vertical, unipolar magnetic field 
of a flux density of 0.001~T (10~G) superposed on a previously computed, relaxed 
model of thermal convection. This flux density is thought to mimic magnetoconvection in 
a very quiet network-cell interior. 
\linebreak The magnetic field is constrained to have vanishing
horizontal components at the top and bottom boundary but lines of force
can freely move in the horizontal direction, allowing for flux concentrations
to extend right to the boundaries. Although this condition is quite stringent, 
especially at the top boundary, it still allows the magnetic field to freely 
expand with height through the photospheric layers into the more or less
homogeneous chromospheric field, different from conventional
simulations that extend to a height of typically 600~km, only.

Subsequent to superposition of the magnetic field, flux expulsion from the granule
centers takes place and within less than 5 minutes,
the magnetic field concentrates in narrow sheets and small knots near the 
surface of optical depth unity with field strengths up to approximately 1~kG. 
Occasionally, these magnetic flux concentrations extend down to the bottom
boundary at a depth of 1400~km but more often, they disperse again at 
a depth of less than 1000~km leaving flux concentrations of a strength 
of a few hundred Gauss only.

\begin{figure*}
  \includegraphics[width=0.98\textwidth]{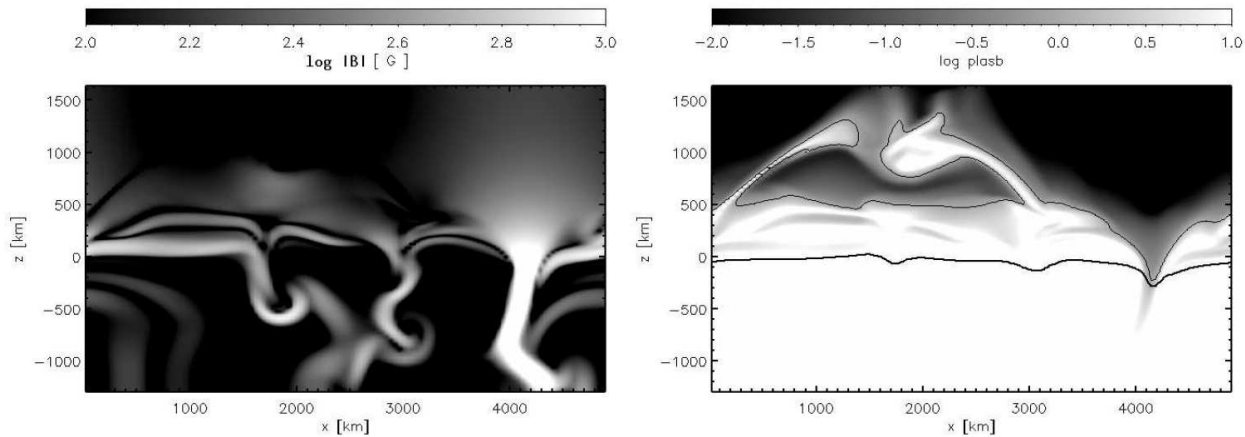}
  \caption{Still images from a time series for the instant $t=1368$ s after starting 
  with an initial homogeneous vertical field of 100~G. {\bf Left:} Logarithmic
  magnetic flux density. {\bf Right:} Logarithm of the ratio of thermal 
  to magnetic energy density (plasma-$\beta$). Also shown are the contour
  of $\beta = 1$ (thin curve) and the surface of optical depth unity,
  $\tau_{500\, \rm nm} = 1$. A strong magnetic flux sheet has formed near
  $x = 4000$~km that causes the funnel of low $\beta$ visible in the 
  right hand panel.
  }
  \label{fig_100G}
\end{figure*}

Fig.~\ref{fig_3Dsection} gives a synoptic
summary of phenomena that occur in this simulation. 
The box on the left hand side renders the logarithm of the magnetic
flux density on a color scale, where red signifies strong (greater or
approximately 0.1 T) and blue weak fields. The surface of equipartition
between magnetic and thermal energy density ($\beta = 1$) is indicated 
as blue surface and on the front side by a yellow curve. It is corrugated 
and continuously changes shape on a time-scale of a few minutes. The black 
curve indicates
optical depth unity. One can see that close to the front side boundary,
a magnetic flux concentration has formed. It leads to a small dip
in the surface of optical depth unity due to the reduced density
within the flux concentration compared to the density in the
environment at the same geometrical height. The box on the right hand
side shows the temperature in colors. We see granules (yellow to light red)
near the height that corresponds to optical depth unity and a network
of cool (dark red) intergranular downflows. In the chromosphere
exists a dynamic, mesh-work like pattern of hot shock waves with cool 
post-shock regions in between.

%%%%%%%%%%%%%%%%%%%%%%%%%%%%%%%%%%%%%%%%%%%%%%%%%%%%%%%%%%%%%%%%%%%%%%%%%%%%%

\section{Local helioseismic experiments with high frequency waves}

Classical helioseismic tools usually rely on the analysis of waves with frequencies 
below the acoustic cutoff that are essentially trapped in the internal 
acoustic cavity of the Sun. These waves are largely evanescent in the outer atmosphere 
and carry little information on this region. The situation changes for waves with 
frequencies above the acoustic cutoff frequency that can freely propagate in the 
atmosphere. When they encounter a change in the dispersive characteristics of the 
medium, they are refracted and reflected (Finsterle et al.~2004).

This effect has been employed by Finsterle et al.~(2004) in order to obtain the 
three-dimensional topography of the `magnetic canopy' in and around active regions 
by determining the propagation behavior of high-frequency acoustic \linebreak
waves in the solar chromosphere. Their method bears 
considerable potential for the exploration of the magnetic field in the solar atmosphere 
by helioseismological means.

In order to test the reliability and to further explore and assess this method we have
carried out a first series of two-dimensional MHD-simulations of the propagation 
of high frequency waves through a magnetically structured, realistic atmosphere. 
We start with a simulation very similar to that described in the previous section,
but carried out in a two-dimensional computational domain of 4900~km
width and spanning a height range of 2900~km, of which 1300~km reach into the 
convection zone and 1600~km above the average level of optical depth unity. 

Since we start with a mean magnetic flux density of 10~G and 100~G, the magnetic 
structures that form in the course of time represent magnetic fields in a very 
quiet network cell interior and in the magnetic network, respectively, rather
than the active region fields observed by Finsterle et al.~(2004). Also, different 
from active region fields, this field evolves on the much shorter time scale of 
granular evolution. Therefore, and for reasons of computational costs, we do not 
analyze a long duration time series as is normally necessary in local helioseismology. 
Instead, we introduce a velocity perturbation of given frequency and amplitude at the 
bottom of the computational domain with which we generate a monochromatic, plane 
parallel wave that propagates within a time span of about 200~s across the height 
range of the computational domain. Experiments with various amplitudes and frequencies 
have been carried out. In the following, however, we only use the values
stated in connection with Eq.~(\ref{eq_sinpert}).

We then determine the arrival time of the wave front at three different heights
in the atmosphere that should roughly represent the heights of maximal Doppler response
of the spectral lines Ni~I 676.8~nm at 200~km, K~I 769.9~nm at 420~km, 
and Na~I 589.0~nm at 800~km (Finsterle et al.~2004). 
Subsequently, we compute the travel times between these heights. They
are thought to become modified through the presence of a strong magnetic field
in the sense that a strong magnetic field reduces the travel time.
At this stage we do not carry out radiation transfer for computing the response of
spectral lines to the perturbation but simply determine the velocity at the
corresponding (fixed) line-formation heights in the computational box for determining
the arrival time of the wave front. The wave front is detected when the velocity
perturbation surpasses a height dependent velocity threshold of less than 100 m/s.

\begin{figure}
  \includegraphics[width=0.48\textwidth]{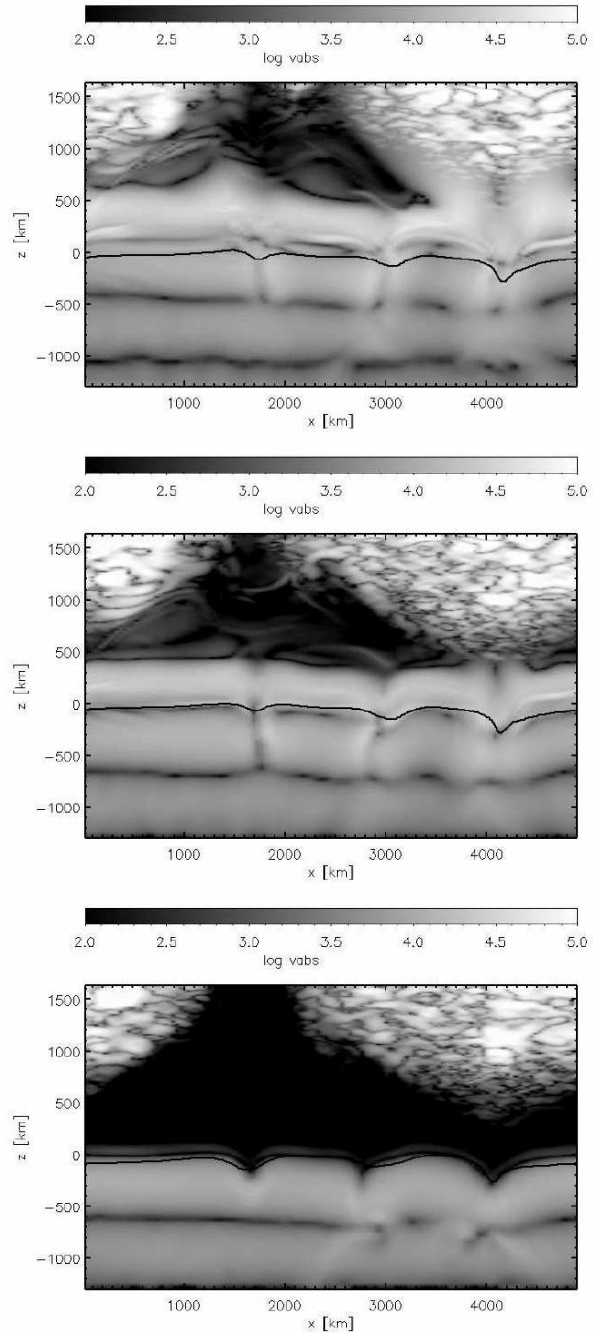}
  \caption{A plane parallel wave with frequency 20 mHz travels through convecting 
   plasma into the magnetically structured photosphere and further into the
   low $\beta$ (magnetically dominated) chromosphere. The three panels
   show the difference in absolute velocity between the perturbed and the
   unperturbed solution 100~s, 148~s, and 168~s (from bottom to top)
   after the start of the perturbation (Eq.~\ref{eq_sinpert}).
   Magnetic field and plasma $\beta$ corresponding to the instant of
   the top panel are given in Fig.~\ref{fig_100G}. A spurious velocity
   signal that is believed to have numerical origin dominates the very
   low $\beta$ region in the upper right and upper left corners. The horizontally
   running black curve near $z=0$ indicates optical depth $\tau_{500\,\rm nm} = 1$.
   The velocity scaling is logarithmic with dimension [cm/s].
  }
  \label{fig_100G20mHz}
\end{figure}

In the following we consider the time span from $t=1200$~s to $t=1400$~s of a 
simulation that started with a homogeneous vertical field of 100~G at $t=0$~s. 
Fig.~\ref{fig_100G} shows a snapshot of the instant $t=1368$~s. 
The panel on the left
hand side shows the logarithmic magnetic flux density, where the values in Gauss
are indicated in the grey-scale bar in the top margin of the figure. A strong
magnetic flux sheet has formed near $x=4000$~km. It leads to a dip in the
surface of optical depth unity, visible in the panel on the right hand side.
There, also the contour of $\beta = 1$, where $\beta = p_{\rm gas}/(B^2/8\pi)$, i.e.,
the contour of equipartition of thermal and magnetic energy density,
is indicated. The magnetic flux concentration causes a funnel of low $\beta$.

The $\beta = 1$ contour delimits the transition from the weak to the
strong field regime. In the latter regime the magnetic field dominates over
thermal pressure. The gas pressure exponentially decreases with height
in a gravitationally stratified atmosphere but the magnetic pressure can
only decrease to a lower limit set by the initial average flux density. Once
the magnetic field, which may be concentrated in small tubes and sheets in
the deep photosphere, has expanded to completely fill the available space,
it cannot further expand and its strength remains essentially constant with 
height. Therefore, $\beta$ unabatedly decreases with height so that with increasing 
height the atmosphere becomes \linebreak
sooner or later magnetically dominated. Correspondingly, 
the $\beta = 1$ contour in Fig.~\ref{fig_100G} (right) extends in a more or
less horizontal direction at a height of 500~km with the exceptions of the
location of the strong flux sheet, where it dips even below $z = 0$, and
two islands higher up in the atmosphere, caused by chromospheric
shock waves.

The $\beta = 1$ contour also marks the region of wave transmission and conversion
(Rosenthal et al.~2002; Bogdan et al.~2003; Cally 2007). For example an
incident acoustic \linebreak
plane wave converts to a fast magnetic wave and gets 
reflected back into the atmosphere. This effect is visible in 
Fig.~\ref{fig_100G20mHz}, which shows, from bottom to top, a time sequence 
of a plane wave traveling through the magnetically structured and
simultaneously evolving atmosphere of Fig.~\ref{fig_100G}.
In order to visualize the wave we have run the simulation twice from 
$t=1200$~s to $t=1400$~s, once without perturbation and once with a
velocity perturbation at the bottom of the computational domain of the form
\begin{equation}
  v_z (t) = v_0 \sin(2\pi(t - t_0)\nu)\,,
  \label{eq_sinpert}
\end{equation}
where $v_0$ was chosen to be 0.05 km/s and $\nu = 20$ mHz. The
two runs were carried out with identical time stepping. Following that,
the two velocity fields are subtracted, which then reveals the wave
perturbation that travels on top of the non-stationary evolving convective 
motion.

Due to a numerical problem of unknown origin, considerable spurious velocities 
occur in the region of very low plasma-$\beta$ in the chromospheric layers 
of the magnetic flux concentration.
This problem will be addressed in a next sta\-ge of the
present project. Nevertheless, we can clearly observe the following phenomena. 
Initially the wave front is retarded at the location of the flux concentration 
(at $x\approx 4000$~km) because of the lower temperature of the plasma within 
the flux concentration, hence the lower sound speed (bottom panel, 100~s after 
start of the perturbation). It undergoes acceleration when entering the funnel 
of $\beta < 1$ so that the wave within the funnel is now leading 148~s after 
start of the perturbation (middle panel).
At the same time the wave becomes fast magnetic in character and starts to
refract. Thus, the wave front becomes inclined until aligned with the
vertical direction and it further turns until traveling into the downward 
direction. In the top panel, only 20 s later, the wave front in the low-$\beta$ 
funnel (extending from $(x,z) = (2400,1500)$ to $(3400,500)$)
has already complete\-ly turned around and travels back {\it into} the 
atmosphere. The\-refore, we can speak of a reflection of the magnetic wave in
the low-$\beta$ region. A similar fanning out of the wave front occurs around 
$x=1400$~km when it reaches the `canopy'-height where $\beta = 1$ at
about 500~km. 

\begin{figure}
  \includegraphics[width=0.48\textwidth]{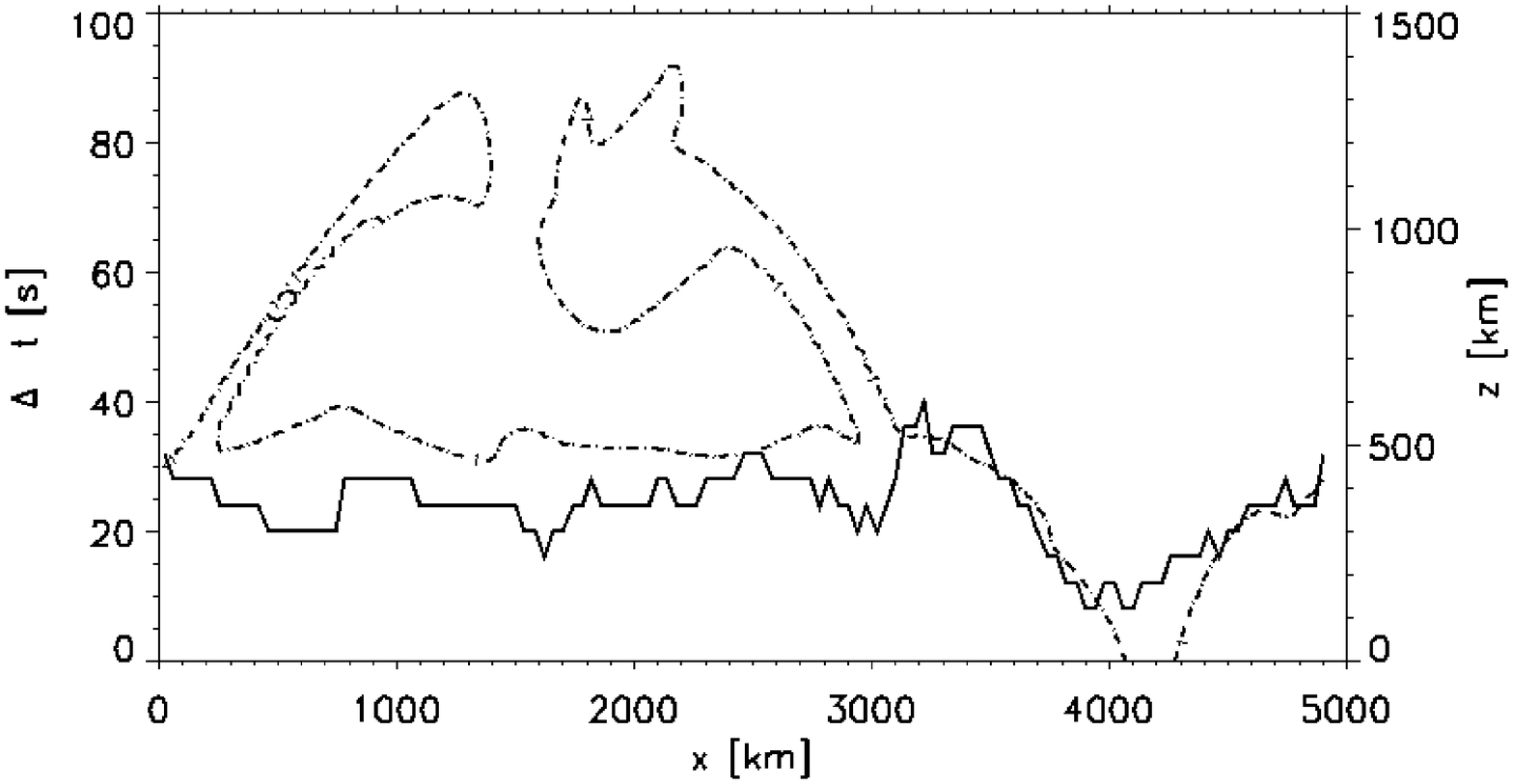}
  \caption{Wave travel time across the layer from $z=200$~km to 
  $z=420$~km as a function of horizontal distance (thick solid curve). 
  Superposed is the contour of $\beta = 1$ (magnetic and thermal equipartition), 
  for which the height is indicated in the right hand side ordinate 
  (dash-dotted curve). Note that the travel time markedly decreases where the 
  low $\beta$ region intrudes this layer.
  }
  \label{fig_traveltime}
\end{figure}

Clearly, the travel time for the wave in the low-$\beta$ region is much 
smaller than elsewhere. Fig.~\ref{fig_traveltime} shows the travel time
between the line formation heights of Ni~I, 676.8~nm at 200~km and
K~I 769.9~nm at 420~km (thick solid curve) as a function of horizontal
distance, $x$. The time axis is given on the left hand side. Superposed
on this plot is the contour of $\beta = 1$ (dash-dotted curve), where the
height in the atmosphere is given by the $z$-axis on the right hand side. 
Clearly, the travel-time curve follows the dip of the $\beta$-contour at 
the location of the flux concentration, which demonstrates that a mapping
of the $\beta = 1$ surface by helioseismic methods is in principle possible.

Because of the spurious velocity noise in the region of very low $\beta$,
we were unable to reliably determine the wave-front arrival at the
line-formation height of Na~I~D2, 589.0 nm at 800~km. But we expect
for the travel time difference between 800 and 420~km an even more 
pronounced dip at the location of the magnetic flux concentration
than for the difference between 200  and 420~km.

We have also carried out a similar simulation with an initial vertical
field of 10~G flux density. Because of the lower field strength the
average height where $\beta = 1$ (canopy height) is in this case at
approximately 1000~km compared to 500~km in the case of 100~G. We
could think of the two simulations with 10~G and 100~G initial field
strength representing very quiet network interior and network regions,
respectively. Examination of the travel times for 20 mHz waves reveals
no major difference in the layer from 200~km to 420~km. In the 100~G case
the mean travel time is 23.9~s and 25.7~s when excluding the strong
flux concentration that has formed at $x\approx 4000$~km. The corresponding
time in the 10~G case is 28.0~s. Since this layer is {\it below} the canopy 
in both cases, there is only a small difference in wave travel-time. However, 
for the layer from 420~km to 800~km we obtain in the 100~G case a travel 
time of 25.1~s (when excluding the strong flux concentration, where reliable
values are unavailable). The corresponding travel time in the 10~G case is 
39.2~s, much longer because in this case this layer is still below the 
low-$\beta$ regime while it is entirely
inside it in the 100~G case. Again this demonstrates that measurements
of the wave travel-time can differentiate between network 
and internetwork regions.

%%%%%%%%%%%%%%%%%%%%%%%%%%%%%%%%%%%%%%%%%%%%%%%%%%%%%%%%%%%%%%%%%%%%%%%%%%%%%

\section{Conclusions}
We have carried out local helioseismic experiments using the 
magnetohydrodynamic code \cobold . A plane-paral\-lel monochromatic wave 
perturbation that is generated at the bottom of the computational domain 
within the convection zone propagates into a non-stationary, 
realistic atmosphere that extends up into the middle chromosphere.
There, it enters a regime in which the magnetic energy density surpasses the 
thermal energy density and converts from acoustic to magnetic in nature.
It is experimentally demonstrated how a magnetic flux concentration leads
to refraction and reflection of the wave.

The experiments also reveal a numerical problem of unknown origin. Two
runs that are identical with the exception of the wave perturbation at the 
bottom of the computational domain show considerable differences in the 
velocity field in regions of $\beta << 10^{-2}$ from the very beginning. In 
these tenuous regions, smallest pressure imbalances that may occur because
of the thermal energy density being only a small fraction of the total energy
density there, must lead to large accelerations. 

We show that the wave travel-time between two fixed levels in the atmosphere
bears information on the nature of the wave and consequently on the magnetic
field. The travel time is reduced in regions of low $\beta$ (strong magnetic
field). For a monochromatic wave of frequency 20 mHz we demonstrate with
Fig.~\ref{fig_traveltime} that the travel time between the heights of 200 and 
420~km in seconds times 15 matches approximately the height of the $\beta = 1$ 
surface in km. In particular, the travel-time curve delineates a funnel
of low $\beta$ that is caused by a local magnetic flux concentration.

The region and magnetic structure considered in the pre\-sent simulation
is much smaller than the active region observed by Finsterle et al.~(2004)
and we use 20 mHz waves, instead of the 7 mHz employed by Finsterle et al.~(2004).
Also, by using plane parallel waves we assume the wave coherence to be always
larger than the magnetic structure under investigation. Subject to these
reservations, the numerical experiments carried out in this 
paper support the conclusions of Finsterle et al.~(2004) and their
proposition using high frequency waves for mapping the magnetic topography 
in the chromosphere. 

\acknowledgements
The authors are very grateful to M.~Haber\-rei\-ter and R.~Hammer for detailed
comments on a draft of this paper.
This work was supported by the {\it Deutsche 
Forschungsgemeinschaft} (DFG), grant Ste 615/5, the German Academic
Exchange Service (DAAD), grant D/05/57687, and the Indian Department of
Science \& Technology (DST), grant DST/INT/DAAD/P146/2006. 
Participation of O. S. at the HELAS workshop in Nice was\linebreak 
supported by the 
European Helio- and Asteroeismology Network (HELAS), which is funded by the 
European Union's Sixth Framework Programme.

\end{document}